\pdfoutput=1
\documentclass{article}

\usepackage{microtype}
\usepackage{graphicx}
\usepackage{subcaption}
\usepackage{booktabs}
\usepackage{hyperref}

\usepackage[preprint]{icml2027}

\usepackage{amsmath}
\usepackage{amssymb}
\usepackage{mathtools}
\usepackage{amsthm}
\usepackage[capitalize,noabbrev]{cleveref}

\theoremstyle{plain}
\newtheorem{proposition}{Proposition}

\icmltitlerunning{Before the Choice: Prospective Representations in Language Models}

\begin{document}

\twocolumn[
  \icmltitle{Before the Choice: An Evidential Framework for Studying\\
    Prospective Representations in Language Models}

  \begin{icmlauthorlist}
    \icmlauthor{Sion Israel Sion}{ets}
  \end{icmlauthorlist}

  \icmlaffiliation{ets}{Department of Software and IT Engineering, \'{E}cole de technologie sup\'{e}rieure (\'{E}TS), Montr\'{e}al, Qu\'{e}bec, Canada}

  \icmlcorrespondingauthor{Sion Israel Sion}{sion-israel.sion.1@ens.etsmtl.ca}

  \icmlkeywords{prospective representation, prospection, forward sweep, language models, mechanistic interpretability, probing, unchosen continuations, counterfactual identifiability, AI welfare}

  \vskip 0.3in
]

\printAffiliationsAndNotice{}

\begin{abstract}
An observable decision reveals the selected action but does not determine which alternatives were internally represented. Drawing on prospection research and hippocampal forward-sweep observations, we develop \emph{Before the Choice}, an evidence-oriented framework distinguishing future-directed content, prospective computation, and causal guidance. The framework organizes four experimental contrasts and makes their alternative explanations explicit. Its first test case concerns unchosen trajectories in deterministic symbolic environments. A preregistered low-rank probe evaluated once on a sealed TEST did not meet its criterion (paired accuracy 0.020 versus 0.195 for a text-and-observed-choice baseline; 293 admissible pairs). Exploratory H1b exposed a reconstruction-by-elimination problem: identifying the chosen candidate can determine the unchosen target. In H1c, matched worlds varied an unchosen continuation while keeping candidate keys bitwise invariant. With specified positive controls passed, decoding in the late-key condition remained at 0.521 ($p = 0.18$; 141 contributing families). These observations do not identify prospective simulation or establish an absence of prospective representations. They illustrate why identifiability, sensitivity, and a dependence-aware null model must constrain inferences from decodability. PREM is proposed as an organizational framework and is not validated by these results. We offer it as a methodological contribution to empirical AI welfare research, which will require such evidential distinctions.
Code and protocol records: \url{https://github.com/besionisrael/before-choice}.
\end{abstract}

\section{Introduction: the problem before the choice}\label{sec:1}

At a junction, an observer sees the action an agent ultimately takes, but not the alternatives the agent may have represented beforehand. A left turn may follow deliberate consideration of two paths, a cached response to a familiar stimulus, or a local decision rule. These possibilities can be behaviorally indistinguishable even though they imply different internal computations.

The resulting scientific problem is one of underdetermination: the observable choice does not uniquely identify its representational causes. This problem is especially consequential when investigating prospective cognition, because the relevant object may be precisely a trajectory that was not selected and therefore never appeared in the system's output.

\begin{figure}[t]
\centering
\includegraphics[width=\columnwidth]{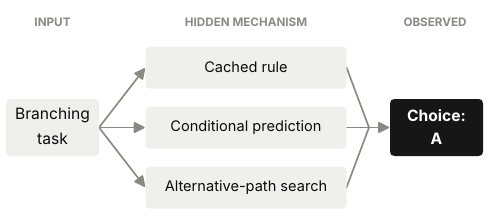}
\caption{Conceptual illustration of behavioral underdetermination. Three hypothetical internal mechanisms produce the same observable action. The figure is not evidence that any tested model uses these mechanisms.}
\label{fig:underdetermination}
\end{figure}

Prospection offers a principled starting point. \citet{seligman2013} describe present representations of possible future states as resources for guiding behavior. \citet{szpunar2014} distinguish simulation, prediction, intention, and planning, preventing the representation of a future possibility from being conflated with any one of these more demanding processes.

Neuroscience provides an influential empirical example. \citet{johnson2007} observed transient hippocampal ensemble activity at rat decision points in which decoded spatial content swept ahead along alternative paths. Their observation motivates the study of future-directed internal representations, but its biological mechanisms cannot simply be assigned to artificial networks.

The analogous question for language models is not whether generated text describes a future. A model may report that an alternative path exists without having simulated it during the current decision. Conversely, internal activations may contain information about future continuations that remains unexpressed. Research on Othello world states \citep{li2023}, future tokens \citep{pal2023}, and unused valid continuations \citep{ustaomeroglu2026} makes the latter possibility empirically plausible while also narrowing any claim of novelty.

The methodological difficulty is that an external probe can decode a target by exploiting information recoverable from the prompt, from the observed choice, or from the candidate representations. Decodability cannot by itself determine whether a model actively constructed an alternative future, nor whether such information influenced the response. Failed decodability likewise has meaning only relative to instrument sensitivity and statistical power.

Before the Choice addresses this evidential gap by separating prospective representational content, prospective computation, and the causal guidance of choice. We formalize the distinction, propose a matrix of discriminating experimental contrasts, and report a first investigation of unchosen trajectories whose negative and exploratory findings illustrate why these inferential boundaries matter.

\section{From decision to prospection}\label{sec:2}

\subsection{Decision mechanisms and future-sensitive behavior}\label{sec:2.1}

A correct action can be generated through learned values, stored policies, explicit transition models, or mixtures of these mechanisms. Model-based reinforcement learning evaluates candidate actions through their consequences, whereas model-free methods can rely on cached values. Successor representations occupy an instructive intermediate position: they store predictions of future state occupancy and can support flexible choices without necessarily simulating every trajectory during each decision \citep{dayan1993,daw2014,momennejad2017}.

Thus, future-sensitive behavior is not sufficient to establish online future simulation. Neither is an internal encoding of alternative outcomes sufficient to establish their use. The target of our research is not decision accuracy alone but the representations and computations that may occur before the response becomes observable.

\subsection{Prospection and the representation of possible futures}\label{sec:2.2}

Prospection concerns present representations of events or states that may arise later. It does not entail retrocausation: the future cannot physically inform a present computation except through information currently available to the system \citep{seligman2013}. In this sense, prospective cognition may operate over action-conditional relations of the form ``if action a occurs, a subsequent state s may follow.''

The taxonomy of \citet{szpunar2014} distinguishes simulation, prediction, intention, and planning and further differentiates episodic from semantic prospection. An abstract symbolic trajectory need not constitute autobiographical imagination. We accordingly use prospective representation as a restricted functional research target within prospection, not as a complete psychological characterization of future-oriented cognition.

\subsection{Memory, construction, and precomputed predictions}\label{sec:2.3}

The prospective brain literature links remembering past events to imagining possible future ones: previously encoded information can be recombined to construct situations that have not occurred \citep{schacter2007}. Computationally, however, it remains essential to distinguish the availability of predictive knowledge from its online construction. A stored environmental model or cached predictive representation can make an alternative accessible without requiring an explicit sweep through that alternative at the point of choice.

In a text-described maze, the full prompt may also provide sufficient information for reconstructing the unchosen route. A decoder trained on hidden states can exploit traces of that structure. The existence of a readable target must therefore be evaluated against explanations involving context, retrieval, and externally learned decoding rules.

\subsection{Deliberation and vicarious trial and error}\label{sec:2.4}

Deliberation is stronger than mere prospective representation: it may involve examining candidate outcomes, evaluating them, and selecting an action through the comparison. Vicarious trial and error (VTE), in which rats pause and orient toward alternatives at maze choice points, has historically served as a behavioral correlate of deliberation \citep{redish2016}. VTE is a behavior, however, not a direct record of conscious thought. An animal can lack overt hesitation while retaining future-relevant information, and an observed pause does not uniquely identify the internal computation.

\subsection{Forward sweeps: observation, interpretation, and transfer limits}\label{sec:2.5}

\citet{johnson2007} recorded hippocampal CA3 ensembles as rats approached demanding choice points. Decoded spatial activity transiently progressed ahead of the rat, sometimes along one route and subsequently another. This observation is more specific than the broad concept of prospection: it involves particular neural populations, spatial coding, and an unfolding neural event.

Related studies have reported hippocampal sequences depicting routes toward remembered goals \citep{pfeiffer2013} and goal-dependent prospective theta sequences \citep{wikenheiser2015}. These are related but physiologically distinct phenomena, not one interchangeable class of ``forward sweeps.'' Together they justify investigating the internal representational content of possible trajectories, while leaving the direction, function, and consciousness of such computation open to separate tests.

\begin{figure*}[t]
\centering
\includegraphics[width=\textwidth]{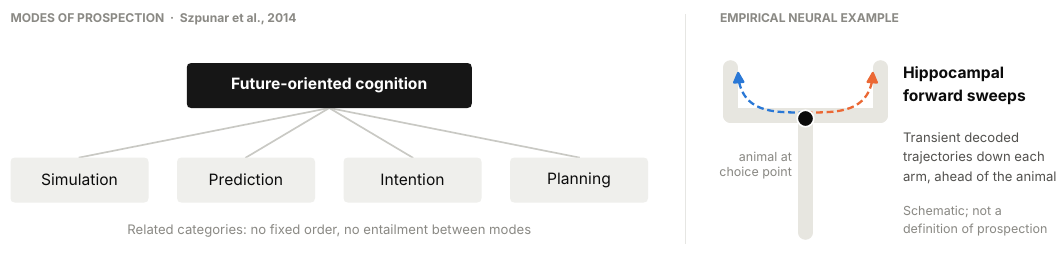}
\caption{Schematic positioning of \citeauthor{szpunar2014}'s four modes of prospection as related categories, not a mandatory sequence. Hippocampal forward sweeps are shown separately as an empirical neural example, not as a necessary biological implementation. No neural recordings are reproduced.}
\label{fig:taxonomy}
\end{figure*}

\subsection{Prediction, alternatives, and counterfactual contrasts}\label{sec:2.6}

Prediction and prospective representation are not opposed: a probability distribution can itself represent several future possibilities. What matters empirically is whether a model encodes action-conditional trajectories, how those relationships are constructed, and whether the model uses them. An unchosen continuation denotes a possible trajectory corresponding to an available but unselected action; it is not necessarily a counterfactual thought held by the model.

We distinguish the researcher's counterfactual manipulation of two task environments from a system's own counterfactual reasoning. Psychological work differentiates episodic future thinking from the construction of alternatives to already realized events \citep{schacter2015}. An experiment can compare worlds in which a path changes without claiming the model itself mentally rehearsed that change.

\subsection{Intentional agency, consciousness, and welfare}\label{sec:2.7}

The capacity to represent conditional future outcomes may support means--end reasoning and, under some accounts, contribute to intentional agency. It is neither necessary for every goal-directed response nor sufficient to establish the broader constellation of goals, beliefs, and desires often associated with intentional agency. Human episodic future thinking can be conscious, but a functional computation representing future consequences need not involve subjective experience \citep{suddendorf2007,dehaene2011}.

AI welfare research similarly distinguishes consciousness, sentience, and forms of agency, and asks separately what would make an artificial system a subject of welfare-relevant interests \citep{long2026}. The present study does not test these stronger properties. Its scope is a minimal, task-relative functional property that could inform later research on agency without being treated as evidence of consciousness, sentience, or moral status.

\section{Prospective representations in language models}\label{sec:3}

\subsection{World-state representations versus future trajectories}\label{sec:3.1}

\citet{li2023} identify a nonlinear representation of Othello board state in a sequence model and use interventions to probe its functional relevance. This demonstrates that sequence models can acquire structured latent representations of an environment. Yet representing a current board state differs conceptually from representing the possible subsequent states under unselected moves.

\subsection{Future tokens and unused continuations}\label{sec:3.2}

\citet{pal2023} show that selected hidden states contain information anticipating tokens beyond the immediate next position. Such generation-relative information differs from the task-relative future states considered here. \citet{ustaomeroglu2026} more directly examine planning horizons and branch awareness, finding that compressed internal representations preserve information about unused correct continuations in several tasks. We therefore make no claim to be the first to investigate unchosen valid paths in language models.

The open inferential problem concerns how one should interpret such signals. A hidden state may expose valid alternatives because it encodes environmental relations, because an online computation assembled their consequences, or because a decoder learned to reconstruct the relevant property. These mechanisms can coexist. Accurately decoding a continuation does not independently identify which one is responsible.

\subsection{Probe selectivity, causal use, and reconstruction}\label{sec:3.3}

\citet{hewitt2019} motivate selectivity controls for probing. \citet{pimentel2020} offer an information-theoretic perspective in which probe performance estimates accessible information; probe simplicity is not, by itself, a sufficient criterion for inference. \citet{ravichander2021} demonstrate that high probing accuracy may arise for properties that are not needed by the model for its task. \citet{elazar2021} distinguish decodable information from behaviorally used information through controlled removal. Causal abstraction provides a formal language for linking interventions to mechanistic claims \citep{geiger2025}. These works supply methodological foundations; PREM does not replace them.

For deterministic models, the complete input constrains subsequent activations. The goal is not to claim hidden states have information-theoretically new knowledge absent from that input; it is to identify how existing information becomes represented, organized, and potentially used by the model. Reconstruction baselines are therefore part of the construct definition, not merely additional performance numbers.

\subsection{Research gap and rationale}\label{sec:3.4}

Existing work already studies future tokens, internal planning, branch awareness, probe selectivity, and causal intervention \citep{pal2023,ustaomeroglu2026,hewitt2019}. Our proposed contribution is narrower than a new probe or the first observation of unused paths: it is an explicit mapping from prospective-cognition claims to discriminating contrasts, including the comparator classes each contrast can exclude. The empirical H1b and H1c cases motivate this mapping through a failed identifiability condition and a repaired one.

\section{The Before the Choice framework}\label{sec:4}

\subsection{Task-relative formulation}\label{sec:4.1}

Let $W$ denote a task environment, $A(W)$ its available actions, and $\tau$ a possible continuation. A task-relative transition structure induces a family or conditional distribution $p(\tau \mid W, a)$ for each candidate action $a$. A model receives $x(W)$, computes an internal state $h$ at a specified measurement point, and emits a choice $a^*$. An unchosen prospective alternative is a continuation associated with an action $a' \neq a^*$. This defines the research target; it does not presuppose that all such continuations are internally represented.

The term ``pre-output'' means recorded before the response is emitted. It does not imply measurement before the model has reached any latent commitment. Likewise, transformer layers index successive transformations of representation, not biological time; a layer profile is not automatically a neural-style temporal sweep.

\subsection{Three analytical dimensions}\label{sec:4.2}

\textbf{Prospective representational content} denotes internal information about possible later task states or trajectories. \textbf{Prospective representational processing} denotes operations that construct, update, transform, or organize such information. \textbf{Prospective guidance of choice} denotes a causal contribution of future-directed information to the final action. These dimensions are related but empirically separable.

\begin{figure*}[t]
\centering
\includegraphics[width=\textwidth]{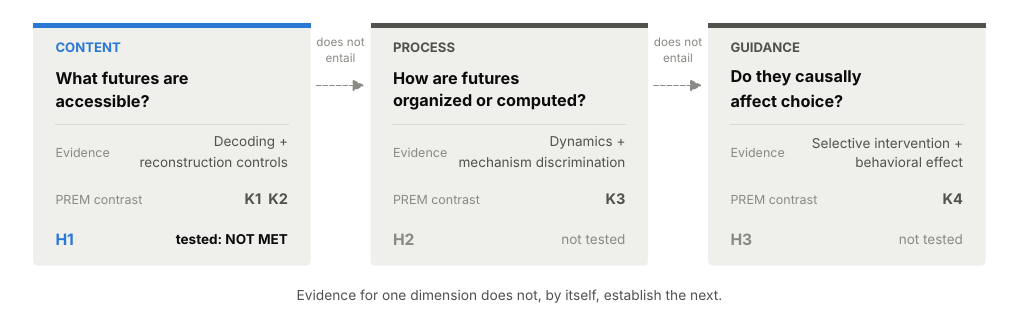}
\caption{The three dimensions of Before the Choice and their corresponding empirical research directions. Dashed arrows mark the absence of entailment: each dimension requires separate evidence. The bottom row gives each hypothesis's status in this manuscript.}
\label{fig:framework}
\end{figure*}

\subsection{Four principles of inference}\label{sec:4.3}

\textbf{P1. Behavioral underdetermination.} A correct choice does not uniquely establish the process that produced it.

\textbf{P2. Content--process non-equivalence.} Decodability of a possible future does not establish an active construction or sweep through that future.

\textbf{P3. Decodability--causality non-equivalence.} Recoverable information need not be functionally used to make the final choice.

\textbf{P4. Detection--absence asymmetry.} Failed detection constrains a particular tested instrument and population; it does not automatically demonstrate absence of the target across the model.

\subsection{Operational requirements}\label{sec:4.4}

We identify five requirements for claims about prospective representational content: (i) a clearly defined future-directed target, (ii) pre-output measurement with an explicit observation site, (iii) sensitivity to manipulated alternatives, (iv) controls against reconstruction from the prompt, choice, and candidate representations, and (v) instrument validation, including positive controls, generalization, and adequate inferential power. Satisfying these requirements would support a narrower content claim; it would not suffice to prove simulation.

\subsection{Prospective Representation Evidence Matrix (PREM)}\label{sec:4.5}

PREM is a provisional map from competing explanations to the evidence capable of constraining them. It is not a new general-purpose probing algorithm or a validated diagnostic instrument. For each contrast, investigators should preregister the target, admissible worlds, which input channels remain invariant, the comparator class excluded by that invariance, the null model, and the strongest permissible claim. K1 and K2 are illustrated by H1b-H1c; K3 and K4 were not investigated in the present experiments.

\begin{table*}[t]
\centering
\caption{PREM contrasts and the limits of their claims.}
\label{tab:1}
\small
\begin{tabular}{p{0.24\textwidth}p{0.39\textwidth}p{0.28\textwidth}}
\toprule
\textbf{Contrast} & \textbf{Manipulation or comparator} & \textbf{Permissible inference} \\
\midrule
K1 --- Alternative specificity & Swap an unchosen continuation while holding choice and other properties as fixed as possible & Signal tracks the alternative under the stated contrast \\
K2 --- Reconstruction resistance & Compare with prompt-only, chosen-action, key-only, and capacity-matched controls & Signal not explained by the tested simpler reconstruction families \\
K3 --- Prospective computation & Manipulate search demands; measure structured internal changes with nuisance controls & Evidence compatible with a particular dynamic or organizational mechanism \\
K4 --- Causal intervention & Perturb a candidate future representation and measure predicted effects & Evidence of causal contribution, conditional on intervention selectivity \\
\bottomrule
\end{tabular}
\end{table*}

\begin{figure*}[t]
\centering
\includegraphics[width=\textwidth]{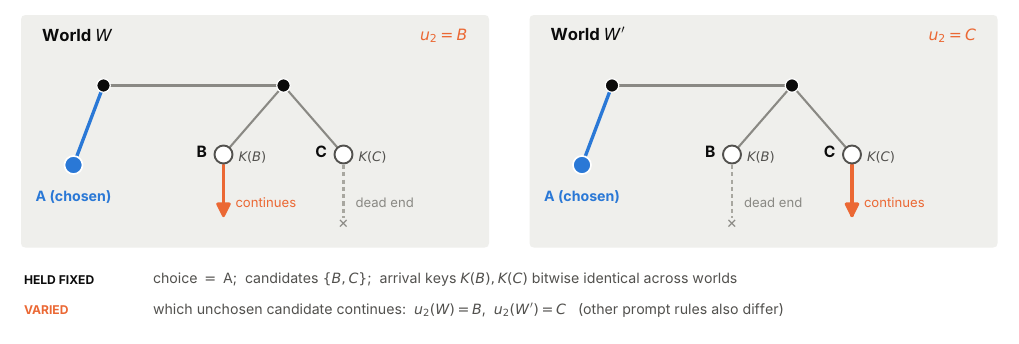}
\caption{Schematic paired-world contrast (not a reproduction of an individual experimental world). Both candidates B and C appear in both worlds; the continuing unchosen candidate changes. The observed choice and candidate labels are held fixed. In the actual H1c late condition, the relevant contextual arrival-key vectors were also verified bitwise identical. Other prompt rules differ between worlds.}
\label{fig:keypair}
\end{figure*}

\subsubsection{A comparator-class identifiability result}\label{sec:4.5.1}

Let $W$ and $W'$ be equally weighted paired worlds, and let $\mathcal{C} = \{B, C\}$ denote the same set of alternative candidates in both worlds. Suppose that the observed action $a^*$ is unchanged and that the contextual arrival-key vectors $K(B)$ and $K(C)$ are bitwise identical across the pair, while the target switches: $u_2(W) = B$ and $u_2(W') = C$. The full prompts necessarily differ in at least one transition rule; no invariance of the entire input is asserted.

\begin{proposition}[Invariant-comparator bound]
\label{prop:bound}
Any deterministic decoder $g$ that receives only $(a^*, \mathcal{C}, K(B), K(C))$, and neither the differing graph rules nor a representation derived from those rules after they change, must return the same candidate in $W$ and $W'$. Since the two correct targets differ, $g$ can be correct in at most one member of each balanced pair. Thus its world-level accuracy cannot exceed $0.5$ on such a balanced pair distribution. The same upper bound holds in expectation for independently randomized decoders whose randomization is not informative about pair orientation.
\end{proposition}

The proposition excludes only the specified information channels; it is not a proof that any successful query-based decoder represents the unchosen future prospectively. The pre-output query may reflect generic relational processing, backward inference from rewards, or other transformations of the differing prompt. The observation is therefore a necessary discriminating contrast for a particular comparator family, not an identification theorem for prospective simulation.

\subsubsection{Dependency-aware null calibration}\label{sec:4.5.2}

A second methodological lesson concerns statistical dependence within a pair. The event that both predictions are correct need not have probability 0.25 merely because each world-level target is binary. Under null prediction strategies that are anticorrelated across pair members, paired success can reach 0.50 without genuine target information. The initial H1c threshold of 0.25 was therefore rejected after null simulations; the revised inferential unit was world-level binary accuracy relative to 0.50, with randomization signs applied to family means. This correction was made before H1c-DEV capture and must not be confused with a positive result for H1c.

A positive K1 contrast demonstrates alternative sensitivity only under the variables and comparator classes actually controlled. Even when $K(B)$ and $K(C)$ are invariant, the full prompt differs, and a model may encode this difference without serial simulation. K2 constrains specified reconstruction strategies, not all nonprospective algorithms. K3 requires a justified process measure; K4 requires selective intervention. Thus PREM specifies which conclusions remain underdetermined after each positive or negative measurement.

\subsection{Research directions and falsifiability}\label{sec:4.6}

H1 examines controlled decodability of unchosen future trajectories from pre-output activations. H2 asks whether internal computations organize or dynamically construct possible trajectories. H3 asks whether candidate representations influence choice under controlled intervention. The three directions are not assumed to form a compulsory developmental sequence; an experiment may be informative about one dimension while remaining agnostic about the others.

Each test should specify a target property, appropriate positive controls, possible alternative explanations, the instrument, its statistical decision rule, and results that would contradict the claim. In particular, an adequately sensitive and powered null observation must be allowed to constrain a specific hypothesis. The framework should not immunize every prospective hypothesis against empirical failure.

\subsection{Claimed contribution and boundary}\label{sec:4.7}

The principal methodological proposal is a record of what each experimental contrast can and cannot exclude. H1b provides a concrete counterexample to unqualified claims of unchosen-trajectory access: a decoder can recover u2 from the chosen depth-2 state. H1c then constructs minimal pairs where the choice and candidate arrival keys remain invariant across worlds while the correct u2 changes. The framework brings together familiar distinctions in a prospective, task-relative setting; its wider utility and novelty over existing probing and causal methods require further comparison and independent applications. The empirical studies below do not validate K3 or K4.

\section{Experimental design and methods}\label{sec:5}

\subsection{Symbolic environments and oracle}\label{sec:5.1}

Turning this question into an experiment requires a situation simple enough that the correct answer can be determined independently of the model, yet rich enough to distinguish competing explanations. Symbolic graphs meet both requirements: they make the alternatives available to the system explicit while allowing certain states to be moved or renamed without altering the optimal decision.

\subsubsection{Graphs, initial choice, and an independent oracle}\label{sec:5.1.1}

Each experimental world is a directed acyclic graph with two initial actions, A and B, two disjoint branches, local transitions, intermediate states, and terminal rewards. A generator controls path depth, node identifiers, and counterfactual variants. An oracle separate from the model computes the optimal action and the states on each trajectory by exact traversal; a second verification implementation checks the consistency of these targets. The complete trajectories and the correct action are not supplied as answers in the prompt sent to the LLM.

In these graphs, A and B do not represent intentions or familiar objects: they simply denote the first two possible actions. This neutrality matters. It helps avoid confusing a preference for learned words with access to decision structure. The oracle defines the correct action and every path\textquotesingle s states; it supplies evaluation targets but no extra answers to the probe at prediction time.

The instruction requests only A or B, without a textual explanation. The forced choice is determined by comparing the logits for the two corresponding tokens and is distinguished from the highest-probability first token over the full vocabulary. This distinction matters when the model prefers an additional word, such as ``To,'' at the beginning of its response: the case is recorded as a formatting violation but is not automatically excluded from the training sample. An exact tie between A and B logits does not define a strict choice and is excluded from admissibility.

A measurement from a single world would remain ambiguous. Counterfactual transformations create related worlds whose relationship is controlled: rename a label, move a rule, change a reward, or alter the abandoned path while tracking what stays constant. The claim is not that a model should ``think like us.'' Rather, a representation genuinely about a particular element should, under suitable conditions, change with that element instead of with incidental cues.

\subsubsection{Counterfactual transformations and controls}\label{sec:5.1.2}

Each family yields 16 worlds: a base world, a counterfactual changing the optimal action, renamings (alias), swaps of action letters (swap), permutations of rule order (shuffle), misleading reward decoys (decoy), changes to the unselected branch preserving the optimal choice (upath), changes to the chosen branch (cpath), and additional state exchanges (xswap). Transformations are applied to both base and counterfactual worlds, yielding eight pairs of variants. Training, validation, and test are split by family rather than by individual row to limit information leakage across variants of the same world.

\begin{table}[t]
\centering
\caption{Controlled graph transformations in H1.}
\label{tab:2}
\small
\begin{tabular}{lp{0.62\columnwidth}}
\toprule
\textbf{Transformation} & \textbf{Main purpose} \\
\midrule
base / cf & Compare outcomes and changes in the optimal action. \\
alias & Test generalization to new identifiers. \\
swap & Control for preference for action letter A or B. \\
shuffle & Control for dependence on rule order. \\
decoy & Explore sensitivity to misleading rewards; outside the H1 threshold. \\
upath & Modify the unchosen branch without altering the optimal choice: primary target. \\
cpath & Measure content associated with the chosen branch separately. \\
xswap & Analyze effects of state identity and presence. \\
\bottomrule
\end{tabular}
\end{table}

\subsubsection{Early revision of the target and backward-reconstruction control}\label{sec:5.1.3}

The first design provided an early methodological warning. A target may appear to concern the future of an action yet in practice be nothing more than a fact printed in the problem statement. Before measuring activations, we therefore checked what a very simple mechanical reader could already recover.

The original target, D-1, was the terminal state of the unchosen branch. The initial audit identified a major conceptual leak: across 300 pilot worlds, the target always corresponded to the minimum reward printed in the prompt. A strategy unable to represent any trajectory could therefore succeed by reading a number. D-1 was removed from the primary analysis before the activation study. Its replacement, D-1$'$, targeted u2, the depth-2 state on the unchosen branch in depth-4 graphs and choice-preserving upath variants.

The u2 target is evaluated among six intermediate-state candidates, ordered canonically by their labels rather than by their ``chosen'' or ``unchosen'' roles. The primary score is the fraction of pairs for which u2 is correct in both worlds. Independent uniform selection from six candidates would yield a paired success rate of $1/36 \approx 0.0278$; this is an intuitive reference point, not a sufficient statistical comparator for H1.

The P0 audit nevertheless showed that a backward reconstructor with access to graph edges can recover unchosen-branch states from a minimum-reward terminal. Depth 4 places u2 farther from a directly visible anchor but does not eliminate this strategy. Any observed decodability must therefore be interpreted as possible relational content, without claiming to identify the direction of the computation that produced it.

Before asking what a model represents during a decision, we must establish that it understands the task well enough to perform it. If its choices are nearly random, a negative decoding result is difficult to interpret. Competence does not imply planning: it simply prevents us from testing for a trajectory representation in a system unable to use the graph rules under the conditions studied.

\subsection{Model selection and datasets}\label{sec:5.2}

\subsubsection{Behavioral competence prerequisite}\label{sec:5.2.1}

The first trial used Qwen3-4B-Instruct-2507. Its preregistered competence checkpoint yielded 9/16 = 56.25\% at depth 2, below the 70\% threshold; performance near chance was also observed at depths 3 and 4. Under the stopping rule, no H1/H2/H3 study was conducted with that model. A new screening protocol then compared three candidates at approximately the 32-billion-parameter scale.

\begin{table*}[t]
\centering
\caption{Model competence screening and selection.}
\label{tab:3}
\small
\begin{tabular}{lll}
\toprule
\textbf{Model} & \textbf{Screening (d2)} & \textbf{Decision} \\
\midrule
Qwen3-4B-Instruct-2507 & 56.25\% (initial checkpoint) & FAIL --- prerequisite not met \\
Qwen3-32B & 61.25\% & SCREEN FAIL \\
OLMo-3.1-32B-Instruct & 55.0\% & SCREEN FAIL \\
Gemma-4-31B-it & 100\% (40/40) & SCREEN PASS; selected \\
\bottomrule
\end{tabular}
\end{table*}

For Gemma-4-31B-it, an independent set of 1,320 worlds met all five competence criteria: 60/60 correct depth-2 decisions; 30/30 depth-2 pairs with both members correct; 1,255/1,320 cases meeting the C3 output-format requirement; 48/60 correct depth-4 answers; and 45/60 admissible depth-4 upath pairs. The format-compliance threshold was passed narrowly, at 0.9508, with 65 initial failures concentrated in decoy worlds. These values establish sufficient competence in the selected conditions, not general planning ability.

\subsubsection{Development and confirmatory datasets}\label{sec:5.2.2}

Two independent datasets were generated using distinct seeds: PROBE-DEV (2026111401) and LOCKED-TEST (2026111402). Each contains 300 families---200 at depth 4 and 100 at depth 3---with 16 variants per family, for 4,800 worlds. Family identifiers, node names, and prompt hashes do not overlap between the two sets. The confirmatory data were procedurally sealed, with authorization and an access log required for release. This seal is a methodological safeguard, not encryption that would resist an administrator.

Within PROBE-DEV, 226 families (3,616 worlds) formed the training subset and 74 families (1,184 worlds) formed the validation subset. After excluding decoys, nonoptimal forced choices, and exact ties, 2,625 training worlds and 855 validation worlds remained; 77 admissible depth-4 upath pairs served as the primary selection set. LOCKED-TEST was held separately and opened only once after sealing, freezing the 16 predictors, and completing the activation capture. Final sample sizes and the verdict are reported in Section~\ref{sec:6.2}.

Once the task has been defined, a less visible question remains: how can we observe an unchosen possibility when the model\textquotesingle s answer contains only one letter? We record an internal state immediately before the answer, then train an auxiliary model to compare that state with the labels of possible candidates. The ability of this auxiliary model---the probe---must be established separately from the LLM\textquotesingle s ability to decide correctly.

\subsection{Activation capture and probing}\label{sec:5.3}

\subsubsection{Observation point and fidelity controls}\label{sec:5.3.1}

The evaluated model is google/gemma-4-31B-it, pinned to revision 842da3794eaa0b77d5f08bae87a17459d91ff475, with bf16 inference, frozen weights, and the official chat template. The measurement is taken at the prompt\textquotesingle s last token immediately before the first answer, at the \textless channel\textbar\textgreater{} position in the selected conversation format. For each world, the pipeline captures 62 vectors of dimension 5,376: the successive API-exposed representations, the final post-normalization state, and a raw vector from the final block before the last normalization.

Activation fidelity is checked by reconstructing logits from pre- and post-normalization states: forced A/B decisions must be preserved, and the maximum allowed deviation over the full vocabulary is 0.25. Near-ties involving the first token under bf16 rounding are logged under technical amendment D3-A2; they do not alter the admissibility rule for strictly tied A/B choices. Each capture includes 20 bitwise replays, finiteness checks, and shard-hash verification before the ephemeral GPU volume is destroyed. The independent LOCKED-TEST capture reproduced these checks, as detailed in Section~\ref{sec:6.2}.

The probe combines two information sources: a query derived from the pre-answer activation, and keys representing the labels from which it must select. These sources must not be conflated. If the keys already contain clues about the path, a good prediction might come from those clues rather than from new information in the query. This distinction, largely hidden in the first version of the instrument, became crucial in H1b and H1c.

\subsubsection{Low-rank bilinear decoding}\label{sec:5.3.2}

The primary probe receives an internal vector $z_\ell$ and a set $C$ of textual candidates. For each candidate c, the vector $e(c)$ is computed from the frozen input embeddings of the tokens in its label, out of context. Scores are estimated through a reduced-rank bilinear form; both Gemma\textquotesingle s parameters and candidate embeddings remain frozen. At inference, the probe receives neither the oracle\textquotesingle s answer, the correct sequence of transitions, nor an indication of which candidate has the role u2.

\begin{equation}
\begin{gathered}
s(c \mid z_\ell) = z_\ell^{\top} U V^{\top} e(c), \\
P(c \mid z_\ell, C) = \frac{\exp(s(c))}{\sum_{j \in C} \exp(s(j))}
\end{gathered}
\end{equation}

Matrices $U$ and $V$ are learned only on PROBE-DEV, with standardization parameters computed on the training split used for selection. Validation uses unseen families; hyperparameters and the number of epochs are fixed before final retraining on all admissible DEV observations. Out-of-context candidate representations were originally selected to prevent the probe\textquotesingle s side input from containing an already-solved graph. Subsequent H1b exploratory diagnostics (Section~\ref{sec:6.3}) exposed the cost of that choice: static keys could not even reliably recover certain labels explicitly printed in the prompt. This finding explains a limitation of the original instrument without retroactively changing the H1 procedure.

\subsubsection{Preregistered search grid and selection}\label{sec:5.3.3}

The representation grid includes 17 indices \{0, 4, 8, \ldots, 56, 60, 61\}, two ranks \{8, 32\}, four regularization coefficients \{$10^{-4}$, $10^{-3}$, $10^{-2}$, $10^{-1}$\}, and three seeds \{0, 1, 2\}. AdamW uses a learning rate of $10^{-3}$, batch size 256, at most 60 epochs, and early stopping with patience 5. The primary selection metric is mean paired accuracy across three seeds on the 77 admissible DEV-validation pairs. Ties favor lower rank, stronger regularization, and then the lower vector index.

The program comprises 408 main fits, 1,632 fits for the four B6 anchor heads, fits for B1 and B5, order heads, B2, final retraining, and 20 conditional logistic-regression fits---2,130 bilinear fits and 20 logistic fits overall. A single counter limits computation to 48 CPU-hours and prohibits selection from an incomplete grid. The 16 final predictors include the three B2 models corresponding to family-level permutations of targets.

\subsubsection{Competing baselines}\label{sec:5.3.4}

\begin{table*}[t]
\centering
\caption{Baseline decoders for the confirmatory H1 test.}
\label{tab:4}
\small
\begin{tabular}{lp{0.5\textwidth}p{0.3\textwidth}}
\toprule
\textbf{Baseline} & \textbf{Construction} & \textbf{Status} \\
\midrule
B1 & Probe using the mean embeddings of prompt positions. & H1 threshold: lexical availability \\
B2 & u2 targets deterministically permuted by family, three permutations; maximum assessed in each bootstrap. & H1 threshold: capacity and memorization \\
B3 & Seven text features + model A/B choice; no path tracing. & H1 threshold: text shortcut \\
B6 & c1/c3/u1/u3 probes; joint assignment of four distinct anchors, followed by two residual candidates. & H1 threshold: anchoring and elimination \\
B4 / B5 / B7 & Answer/reward decoder; candidate-only model; chosen branch cpath. & Descriptive; outside H1 threshold \\
\bottomrule
\end{tabular}
\end{table*}

For B6, joint assignment of anchors leaves two distinct candidates; the expected probability of recovering u2 in a world is 1/2 when its true target is among those two, and zero otherwise. Across a pair, the expectation is the product of the two world-level probabilities, hence either 0 or 1/4. These figures are not the observed accuracy of the B6 heads themselves. At depth 3, B2 permutes the four existing positions rather than six, without artificially mapping back to the correct target or requiring a derangement.

\subsubsection{H1 statistical decision and L2 attribution}\label{sec:5.3.5}

The confirmatory verdict applies exclusively to LOCKED-TEST. Before any comparison, the result is declared INSUFFICIENT if fewer than 180 depth-4 upath pairs are admissible or fewer than 100 families contribute. Otherwise, for each of the four baselines $k \in \{\mathrm{B1}, \mathrm{B2}, \mathrm{B3}, \mathrm{B6}\}$, we estimate $\Delta_k$, the difference between probe and baseline paired accuracy. A 95\% percentile bootstrap with 10,000 replications resamples families. For B2, the maximum baseline across its three permutations is recalculated within each replicate. H1 is MET if and only if all four lower confidence bounds for $\Delta_k$ are strictly positive.

\begin{equation}
\begin{gathered}
\Delta_k = \mathrm{PairedAcc}(\text{probe}) - \mathrm{PairedAcc}(B_k), \\
\text{H1 MET} \iff \min_k \mathrm{LB}_{95\%}(\Delta_k) > 0
\end{gathered}
\end{equation}

A positive H1 would permit an L1 conclusion, but not automatically L2. L2 requires three additional tests, all at the same layer index $\ell^*$ selected for H1: (i) exact joint decoding of the ordered triple u1/u2/u3 above max(1/6, B3-order); (ii) generalization to alias variants; and (iii) generalization to shuffle variants, each above max(1/6, B3), with a positive bootstrap lower bound. None of these criteria establishes forward computation or a causal role in choice.

The first test answers a modest but rigorous question: does our instrument, as frozen before TEST, outperform nonprospective explanations on previously unseen families? It is not designed to distinguish every form of internal computation. DEV results prepare the instrument; they cannot prove H1 or retroactively modify its confirmatory decision rule.

\section{Empirical results}\label{sec:6}

\subsection{H1 development analyses}\label{sec:6.1}

\subsubsection{Model selection and competence}\label{sec:6.1.1}

The initial Qwen3-4B model failed the behavioral prerequisite (56.25\% at the depth-2 checkpoint); two subsequent 32-billion-parameter candidates also failed screening. Gemma-4-31B-it passed screening and then the independent C1--C5 competence check. Its observed depth-2 accuracy (60/60) substantially exceeded that of the other candidates under the conditions studied, whereas depth-4 performance declined to 48/60 = 0.800. This result supports the feasibility of probing representations; it does not demonstrate that alternative trajectories are already represented.

\subsubsection{DEV capture: completeness and numerical quality}\label{sec:6.1.2}

An initial D3-DEV GPU run stopped because of a numerical near-tie for the first token: the maximum reconstruction deviation remained acceptable and the forced A/B decision was unchanged, but the full-vocabulary argmax differed. No activations were retained from this incident. After technical amendments were tested, one authorized D3-DEV-R1 rerun captured all 4,800 worlds on an A100 80 GB. Each world yielded 62 vectors of 5,376 components. All 20 replays were bitwise identical; no nonfinite values were reported; the maximum logit deviations were 0.195 (post-normalization) and 0.205 (pre-normalization), below the 0.25 threshold.

Fourteen worlds were logged for either an original A/B tie or a top-1 near-tie: 11 A/B ties and four near-ties, with one overlap. The 12 data shards, prompt embedding means, and 5,800 candidate embeddings were verified. GPU capture time was reported as 590 seconds, with peak memory of 63.35 GB, 3.43 GB transferred, and marginal cloud cost of US\$0.579. ``Valid capture'' refers to the instrument\textquotesingle s numerical fidelity, not to the validity of H1.

\subsubsection{Training sample, format, and admissibility}\label{sec:6.1.3}

Of the 4,800 DEV worlds, 600 were decoys excluded from training. Among 4,200 nondecoy worlds, 710 forced A/B decisions were incorrect and 10 were ties, leaving 3,480 training-admissible worlds: 2,625 in training and 855 in validation. The overall first token differed from A/B in 471 worlds, including 396 decoys and 75 nondecoys. All 75 nondecoy format violations occurred at depth 4 and began with ``To''; 37 were retained because their forced A/B choice was correct, as required by the frozen rules. The 77 admissible DEV-validation upath pairs determined primary probe selection.

\subsubsection{DEV performance: a negative exploratory result}\label{sec:6.1.4}

The single full CPU run took approximately 35 minutes of wall-clock time and 1.48 CPU-hours. It completed the 2,150 planned fits and stored the 16 final predictors with hashes. No fit reached the 60-epoch cap: early stopping occurred primarily between epochs 3 and 14. The selected primary probe used vector index 32, rank 8, weight decay 0.1, and five epochs, all selected solely on DEV-validation.

\begin{table*}[t]
\centering
\caption{H1 development performance and diagnostics.}
\label{tab:5}
\small
\begin{tabular}{lllp{0.33\textwidth}}
\toprule
\textbf{Measure} & \textbf{DEV validation} & \textbf{DEV training} & \textbf{Interpretation} \\
\midrule
H1, u2 (paired) & 0.030 & 0.071 & At simple chance level ($\approx 0.028$). \\
B1, prompt mean (paired) & 0.000 & NA & Lexical control; no observed advantage. \\
B5, candidate-only (paired) & 0.000 & NA & Descriptive control. \\
B3, text + choice (paired) & 0.156 & 0.268 & Outperforms H1 probe on DEV. \\
Individual B6 heads & 0.24--0.31 & 0.80--0.93 & Head-level scores; NOT the paired B6 score. \\
u1 / u3 order heads & 0.19 / 0.17 & 0.4--0.6 & Limited generalization. \\
B3-order u1 / u2 / u3 & 1.000 / 0.327 / 0.222 & NA & u1 is readable from the action line. \\
\bottomrule
\end{tabular}
\end{table*}

The best primary configuration at any layer, averaged over three seeds, did not exceed 0.030 paired accuracy on DEV-validation. Under this architecture and selection procedure, the probe thus provided no convincing relational signal. B3, by contrast, reached 0.156: simple textual features produced a substantially stronger baseline. The training--validation gaps for B6 anchors and order heads point to considerable difficulty generalizing to unseen families, but do not identify a single cause.

The exploratory results document instrument selection without determining the H1 verdict. Weights, selected configurations, normalization statistics, training logs, and hashes of the 16 predictors are referenced in the BTC-H1 v0.3 research archive; the protocol and evaluator were frozen before analytical access to LOCKED-TEST. Unfavorable DEV observations did not trigger any change to the confirmatory criterion.

Weak performance on development families could have motivated changing the layer, the score, or the candidate representation. Precisely this is forbidden by the experimental freeze at confirmation time. The independent TEST must evaluate the method selected on DEV, even if its likely result appears unfavorable. In this way, we separate the question ``How well did this preregistered hypothesis perform?'' from the later question ``Why did our first measurement fail?''

\subsection{H1 confirmatory evaluation}\label{sec:6.2}

After freezing the H1 v0.3 protocol and its 16 predictors, the model was run on all 4,800 LOCKED-TEST worlds. Activations were sealed separately from oracle files, then evaluated in a single analytical access after hash verification and creation of the EVAL\_STARTED marker. Confirmatory computation ran on CPU for 30 seconds using four threads; no retraining, hyperparameter selection, or threshold adjustment occurred after release.

\subsubsection{TEST capture and activation integrity}\label{sec:6.2.1}

The first TEST capture attempt failed during container startup, before any forward pass, because of an incorrect local path; it produced neither activations nor analysis and was stopped and archived. The authorized BTC-D3-TEST-R1 rerun succeeded on an A100 80 GB: 4,800/4,800 worlds, 12 shards, 62 vectors of 5,376 dimensions per world, approximately 3.43 GB transferred, and 5,800 verified candidate embeddings. The run lasted 17 min 27 s, with peak GPU memory of 63.35 GB and cloud cost of US\$0.746. All 20 replays were bitwise identical; no nonfinite values or fidelity errors occurred. The maximum reconstructed-logit deviation was 0.1952, below the 0.25 threshold. Nine worlds exhibited exact A/B ties and seven others exhibited first-token near-ties, logged under D3-A2.

\subsubsection{Confirmatory population and admissibility}\label{sec:6.2.2}

Admissible worlds are those for which the model\textquotesingle s forced A/B choice is strictly defined and matches the oracle\textquotesingle s optimal action. Of the 4,800 TEST worlds, nine exact ties and 1,000 nonoptimal choices were excluded, leaving 3,791 admissible worlds. Among these, 359 decoys could not be evaluated by the frozen six-candidate decoders, yielding 3,432 evaluable worlds. The primary endpoint, however, uses only depth-4 upath pairs: 293 pairs from 200 families, above both preregistered sufficiency thresholds (180 pairs and 100 families).

\begin{table}[t]
\centering
\caption{H1 confirmatory population and admissibility on LOCKED-TEST. Decoys have 10 or 7 candidates versus 6 for the frozen decoders. Sufficiency required $\geq$180 depth-4 upath pairs and $\geq$100 families.}
\label{tab:6}
\footnotesize
\begin{tabular}{lr}
\toprule
\textbf{Confirmatory indicator} & \textbf{LOCKED-TEST value} \\
\midrule
Generated worlds / families & 4,800 / 300 \\
Exact A/B ties excluded & 9 worlds \\
Nonoptimal choices excluded & 1,000 worlds \\
Admissible worlds & 3,791 \\
Admissible decoys, not evaluable & 359 \\
Evaluable worlds & 3,432 \\
Admissible d4 upath pairs & 293 \\
Contributing families & 200 \\
Preregistered sufficiency & MET \\
\bottomrule
\end{tabular}
\end{table}

\subsubsection{Primary result: four preregistered contrasts}\label{sec:6.2.3}

On the primary population of 293 pairs, the H1 probe (index $\ell^* = 32$, rank 8, regularization 0.1, five epochs) achieved paired accuracy 0.020, with a 95\% bootstrap interval of {[}0.007, 0.038{]}. The theoretical chance rate for independent predictions among six candidates in both worlds is $1/36 \approx 0.028$. The baselines obtained B1 = 0.000, B2$_{\max}$ = 0.020, B3 = 0.195, and B6 $\approx$ 0.020. The intersection--union test uses 10,000 family-level resamples (seed 0) and requires four strictly positive lower bounds for the accuracy differences.

\begin{table*}[t]
\centering
\caption{H1 confirmatory result: the four preregistered contrasts.}
\label{tab:7}
\small
\begin{tabular}{lrrc}
\toprule
\textbf{Contrast (baseline acc.)} & \textbf{$\Delta$ probe $-$ baseline} & \textbf{95\% bootstrap CI} & \textbf{Lower bound $>0$?} \\
\midrule
H1 $-$ B1 (0.000) & $+0.020$ & $[+0.007, +0.038]$ & Yes \\
H1 $-$ B2$_{\max}$ (0.020) & $0.000$ & $[-0.021, +0.020]$ & No \\
H1 $-$ B3 (0.195) & $-0.174$ & $[-0.220, -0.129]$ & No \\
H1 $-$ B6 ($\approx 0.020$) & $+0.001$ & $[-0.016, +0.020]$ & No \\
\bottomrule
\end{tabular}
\end{table*}

\begin{figure*}[t]
\centering
\includegraphics[width=\textwidth]{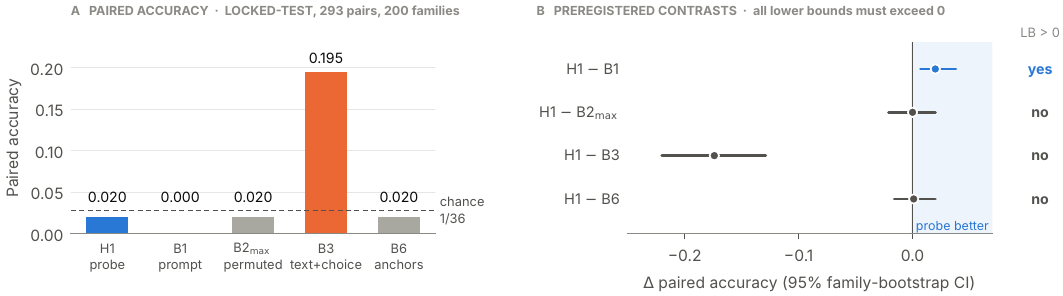}
\caption{Confirmatory H1 paired accuracy. B3 uses input-text features plus the model\textquotesingle s observed A/B choice; it is not a prompt-only baseline. (a) The dashed 1/36 line is descriptive chance. (b) Preregistered success required the 95\% family-bootstrap lower bound of every contrast to exceed zero; only the contrast with B1 did. H1 NOT MET.}
\label{fig:h1}
\end{figure*}

The comparison with B1 passed, but those against B2, B3, and B6 failed. In particular, the difference from the text-and-observed-choice baseline B3 was negative and its confidence interval excluded zero: $\Delta = -0.174$ $[-0.220, -0.129]$. Because H1 required all four contrasts to pass simultaneously, the preregistered verdict is H1 NOT MET. This verdict denotes failure to meet the criterion for decodability beyond the baselines; it does not demonstrate that the information is absent from every possible representation.

\subsubsection{Conditional L2 attribution}\label{sec:6.2.4}

Under the preregistered plan, ordered-triplet u1/u2/u3 tests and alias/shuffle generalization tests for L2 attribution are triggered only if H1 is MET. Since that condition was not satisfied, L2 was not evaluated. No L1 or L2 evidence level is assigned in this study. H2 (descriptive organization of trajectories) and H3 (causal contribution) were not tested.

\subsubsection{Secondary analyses and descriptive observations}\label{sec:6.2.5}

Secondary analyses were preregistered with different statuses. Inferential tests S1 and S2 received Holm adjustment, applied only within this subset of the secondary plan. The remaining analyses were descriptive, infeasible with frozen predictors, or not performed.

\begin{table*}[t]
\centering
\caption{H1 secondary analyses and descriptive observations.}
\label{tab:8}
\small
\begin{tabular}{llp{0.55\textwidth}}
\toprule
\textbf{Analysis} & \textbf{Status} & \textbf{TEST result} \\
\midrule
S1: upath d3+d4 & Inferential, Holm & 467 pairs / 300 families; H1 0.015, B3 0.319; $\Delta = -0.304$; adjusted $p = 1.0$. \\
S2: xswap d4 & Inferential, Holm & 286 pairs / 197 families; H1 0.007, B3 0.234; $\Delta = -0.227$; adjusted $p = 1.0$. \\
S3: depth-specific heads & Descriptive & u1 0.191; u2 0.198; u3 0.205; anchors c1 0.188, c3 0.363, u1 0.238, u3 0.274 (chance 0.167). \\
S4: decoys & Not computable & Decoys have 7 or 10 candidates vs six for the frozen decoder. \\
H2: layer profiles & NOT RUN & Would require training new probes at different vector indices. \\
B7: chosen branch & Descriptive & Paired accuracy 0.119 on 286 cpath pairs; outside H1 threshold. \\
B4: answer/rewards & Not computed & Theoretical baseline $= 0$ by construction of D-1$'$ target. \\
\bottomrule
\end{tabular}
\end{table*}

Descriptive anchor scores, especially c3 = 0.363, and B7 = 0.119 are consistent with a possible decodability advantage for the chosen branch. However, no controlled preregistered comparison warrants a claim of such an asymmetry. The adjusted p-values of 1.0 for S1 and S2 arise from one-sided tests seeking superiority of the probe, whereas the observed effect points in the opposite direction.

\subsubsection{Summary of the confirmatory finding}\label{sec:6.2.6}

The confirmatory protocol was not underpowered in terms of its preregistered sample sufficiency criteria: 293 pairs from 200 families contributed to the primary contrast. The development result was qualitatively replicated on independent families: the probe remained near the idealized chance level and far below the text-and-observed-choice baseline, even though the model had passed its behavioral competence prerequisite. H1 v0.3 was therefore closed as NOT MET. Data seals remain intact and accesses are archived; LOCKED-TEST is consumed and cannot be used to select new methods.

At that point, we knew that the H1 probe had failed the announced test. We did not yet know why. Two explanations were compatible with the same result: either the activations did not make the unchosen state accessible under the tested conditions, or the instrument could not read information that was nevertheless present. H1b arose from this distinction; it was not a second attempt to confirm H1 on the already-consumed TEST set.

\subsection{H1b: instrument and identifiability diagnostics}\label{sec:6.3}

The H1 confirmatory result was preserved without reinterpretation: LOCKED-TEST was consumed, and no additional training on it was permitted. A separate investigation, BTC-H1b, then used PROBE-DEV to examine possible explanations for the negative result. Because many configurations were inspected on this same development dataset, all H1b comparisons are exploratory; no H1b confirmatory verdict was issued.

The first check was deliberately simpler than u2 itself. If a probe cannot retrieve a label explicitly associated with the chosen action, its failure on an unchosen trajectory cannot be interpreted as an observation about prospection. We therefore examined separately the extraction of scalar variables and the binding between representations and arbitrary identifiers.

\subsubsection{Scalar variables are readable, but label bindings are weak}\label{sec:6.3.1}

Experiments X1/X2/X7 show that a linear decoder can accurately read several scalar properties from activations, including the model\textquotesingle s choice (approximately 0.97--1.00 depending on layer) and reward gap (approximately 0.93). By contrast, the original bilinear probe, combining an internal query with static mean embeddings of random labels, could not reliably identify even states explicitly present in the prompt, such as c1 or A1. A text-based control readily recovered those references. The H1 result must therefore be stated as a failure of the chosen measurement pipeline, not a demonstration that Gemma lacks any representation of alternative paths.

This investigation shifted the diagnosis: several global properties were easily decodable, but the labels themselves were not accessible with out-of-context keys. Replacing those keys with representations taken when the model reads each label occurrence changes what the probe can learn. It improves measurement, but introduces a new question: do contextual keys already carry enough graph information to reconstruct the target without any additional encoding in the pre-answer state?

\subsubsection{Contextual keys make states decodable}\label{sec:6.3.2}

Experiments X3/X4 captured multiple reading and query positions to represent each candidate by its contextual activation at the arrival occurrence. Static keys remained near chance, whereas contextual keys yielded approximately 0.96 accuracy for c1 and up to 0.82 for u2 in individual worlds; the pointer reached approximately 0.69 paired u2 accuracy in its best explored configuration. These results show that label-associated representations can carry information about branch and depth structure. They do not establish that the decision state contributes information specific to the unchosen continuation.

This qualification is substantive: the ``chosen letter + contextual keys'' control recovered c1 and A1 with accuracy 1.00 in some settings. An initial bilinear control matched for parameter count reached only 0.403 on paired u2, compared with 0.571 for a more competitive linear control; matching parameter counts does not guarantee matching effective capacity. This diagnosis prompted stronger baselines rather than attributing every apparent advantage to the pointer.

\begin{table*}[t]
\centering
\caption{H1b exploratory diagnostics and identifiability.}
\label{tab:9}
\small
\begin{tabular}{p{0.22\textwidth}p{0.3\textwidth}p{0.38\textwidth}}
\toprule
\textbf{Instrument or contrast} & \textbf{Observed performance} & \textbf{Cautious interpretation} \\
\midrule
H1 probe, static keys & c1 / A1 $\approx$ chance & Sensitivity to label identity not established \\
Scalar variables, linear probes & choice $\approx$ 0.97--1.00; reward gap $\approx$ 0.93 & Some information is accessible to probes \\
Contextual arrival keys, X3/X4 & c1 $\approx$ 0.96; u2 $\leq$ 0.82 (single), $\approx$ 0.69 (paired) & Graph information is decodable, without evidence of a query-specific contribution \\
Choice + keys control & c1 / A1 up to 1.00 & Risk of reconstruction without a decision-state query \\
Elimination through c2 & paired u2 0.818 vs pointer 0.688 & The original task cannot distinguish the two information sources \\
\bottomrule
\end{tabular}
\end{table*}

The elimination control makes the problem tangible. In this first task, depth 2 contains only two possible states, one chosen and the other unchosen. Once the chosen state is recognized and both candidate labels are known, the other is immediately determined. A probe can thus appear to read the unchosen trajectory while actually using a clue about the chosen trajectory and a subtraction operation. No decoding accuracy on this target alone can distinguish these two explanations.

\subsubsection{Why H1b could not become a confirmation}\label{sec:6.3.3}

In the H1 task, exactly two labels appear at the second level, c2 and u2. Once c2 and the prompt structure are known, u2 follows by elimination. On DEV, a control that first predicts c2 and then selects the other candidate achieved 0.818 paired accuracy, compared with 0.688 for the u2 pointer. This does not prove that every internal representation of u2 is derived from c2; it does show that decoding accuracy alone cannot isolate information specific to the unchosen branch in this task. The BTC-H1b v1.0 protocol was prepared and hashed but never run on a new TEST; it was suspended before pool generation. The identifiability problem motivated H1c rather than another underspecified test.

The correction required by H1b was therefore more than a better decoder. The experiment itself had to be redesigned so that the target could not be inferred by construction from the choice and a candidate list. This is the idea behind H1c: create two nearly identical worlds in which the chosen branch stays fixed but the state that continues the unchosen branch changes.

\subsection{H1c: controlled minimal pairs and exploratory results}\label{sec:6.4}

The design resembles a swap of roles. Two successor states exist on the abandoned branch. In one world, the first continues toward a reward while the other leads to a dead end; in the paired world, their roles are reversed. The available labels and chosen action remain unchanged. To succeed on both worlds, a probe must alter its prediction in response to the relevant difference rather than repeat a candidate dictated by a constant feature.

\subsubsection{From the original graph to symmetric dead ends}\label{sec:6.4.1}

H1c modifies the two-action graph so that each depth-1 node has two chains of equal length: a continuation reaching a reward and a terminal dead-end chain with no reward. Adding three states to the dead-end chain corrected two frequency-based label shortcuts, successively discovered at depths 2 and 3. The task contains twelve intermediate-state candidates. Within each ``late'' pair (base\_L, dswap\_L), swapping the source labels of two rule lines changes which of the two unchosen continuations reaches the reward, while preserving the optimal choice, c2, the text of the chosen branch, the candidate set, branch membership, and label occurrence counts.

In the late ordering, the two contextual arrival keys queried by the pointer occur before the first changed token. Their causal prefixes and positions are identical across pair members; the corresponding activations were verified to be bitwise identical in 400 DEV pairs and 600 DEVX pairs. Thus, a deterministic decoder seeing only candidate keys and the choice must give the same answer even though the correct continuation changes. In the ``early'' ordering, the modified rules precede the arrival occurrences: the keys can therefore differ, providing an additional exploratory condition. A solver using the complete graph structure and rewards can nevertheless recover the correct continuation in every world; H1c does not establish that this information was computed in a prospective direction.

For this swap to have its intended meaning, matching label lists is not enough. Initially, occurrence counts revealed which path continued; lengthening the dead-end chains removed these cues. Rule order then became an experimental condition: in late pairs, both candidate keys are computed before the differing lines. Their bitwise identity was verified in the activations. In early pairs, those lines are read before the keys, which can then vary. Contrasting these conditions lets us ask what the keys contribute and what remains accessible from the pre-answer query.

A second correction concerned not the graph design but the meaning of ``better than chance.'' When the correct answer switches between members of a pair, the decoder\textquotesingle s two predictions need not be independent. An intuitive paired-accuracy threshold can therefore be misleading even when every graph is generated correctly.

\subsubsection{A necessary correction to statistical inference}\label{sec:6.4.2}

The initial rule counted a pair as successful when both members were predicted correctly, assuming a chance level of 0.25. That value is correct only under particular assumptions about prediction dependence. A null simulation with perfectly anticorrelated prediction preferences but no information about the true orientation produced 100\% false rejections under this rule: predictions were opposite across pair members, and paired success reached 0.50 under the null. Amendment A1, documented and frozen before the H1c-DEV capture, therefore uses per-world binary accuracy against 0.50 and a sign-flip randomization test on family-level means. Pair orientations were audited. Across 5,000 null datasets per scenario, corrected rejection rates were 0.0246 for independent preferences and 0.0224 for anticorrelated preferences, near or below the one-sided 0.025 threshold. This correction illustrates how an apparently natural baseline can invalidate an inference when it ignores dependence structure.

Before treating the absence of a signal as informative, we defined sensitivity controls. Among other things, the instrument had to recognize a simple label linked to the chosen action and identify a chosen continuation when the design allowed that measurement. Failure of the initial positive control prompted a layer and sample-size diagnosis, rather than a premature conclusion about the abandoned branch.

\subsubsection{Initial DEV: integrity passed, instrument insufficient}\label{sec:6.4.3}

The first H1c-DEV capture of 2,000 worlds passed quality gates: state fidelity, 20 bitwise replays, key identity, and behavioral competence of 0.723. Binary decoding of the unchosen continuation in the late condition remained at chance (0.479; p = 0.75). However, positive control PC1, designed to identify c1 among twelve candidates, reached only 0.399 at layer 56, selected by an essentially chance-level binary grid; this fell below the preregistered 0.80 threshold. The unchosen-continuation result was therefore classified as inconclusive because of insufficient instrument sensitivity. A post hoc diagnostic found PC1 = 0.752 at layer 40, still below threshold, with an increasing learning curve and recurrent confusion between c1 and u1.

DEVX addressed a specific question: with more families and a layer selected using a positive control independent of u2, could the probe now perform the elementary readings expected of it? Only after this requirement was met could unchosen-branch performance be described as an interpretable exploratory result for this particular instrument.

\subsubsection{DEVX: independent sensitivity control and results}\label{sec:6.4.4}

A single extension of 3,000 worlds added to the initial DEV data yielded 1,000 families in total, split into TRAIN (700), SELECT (100 new), and EVAL (200 new). Previously analyzed families participated in neither SELECT nor EVAL. Layer selection, independent of u2 decoding, used PC1 on SELECT and chose layer 40. On EVAL, 711 admissible worlds from 174 contributing families were available. All three criteria fixed before capture were passed: PC1 = 0.886 {[}0.855, 0.915{]} (threshold 0.80), SELECT--EVAL gap = 0.023 (threshold 0.10), and chosen-continuation PC2 decoding = 0.756 {[}0.720, 0.795{]} (p \textless{} $10^{-4}$). The preregistered decision was only ``INSTRUMENT ADEQUATE''; it did not establish success for the hypothesis about the unchosen branch.

For late unchosen-continuation pairs, the pointer achieved world-level binary accuracy 0.521 with a reported descriptive 95\% interval {[}0.480, 0.557{]} and a family-level sign-flip p = 0.18 (141 contributing families): no advantage over 0.50 was detected. The descriptive interval and the sign-flip test are distinct summaries and should not be assumed to use the same sampling model. For early pairs, where candidate keys differ across worlds, accuracy reached 0.703 (118 families; family sign-flip $p < 10^{-4}$). This contextual difference does not show that the keys alone carry the information: a query-key interaction remains possible. No H1c TEST pool was generated, and no confirmatory H1c evaluation was conducted.

\begin{table*}[t]
\centering
\caption{H1c exploratory outcomes and sensitivity controls.}
\label{tab:10}
\small
\begin{tabular}{lll}
\toprule
\textbf{Stage / condition} & \textbf{Accuracy} & \textbf{Population / status} \\
\midrule
Initial DEV --- late unchosen & 0.479; $p = 0.75$ & Exploratory; insufficient instrument (PC1 = 0.399) \\
DEVX --- PC1, 12 candidates & 0.886 $[0.855, 0.915]$ & Preregistered sensitivity control; EVAL, 174 families \\
DEVX --- chosen continuation & 0.756 $[0.720, 0.795]$ & Positive control SC3; EVAL, all variants \\
DEVX --- late unchosen & 0.521 $[0.480, 0.557]$; $p = 0.18$ & Exploratory observation; 141 families \\
DEVX --- early unchosen & 0.703 $[0.648, 0.758]$ & Exploratory; 118 families \\
cswap, chosen, invariant keys & 0.712 $[0.596, 0.827]$ & Post hoc; 26 families, restricted subset \\
\bottomrule
\end{tabular}
\end{table*}

\begin{figure}[t]
\centering
\includegraphics[width=\columnwidth]{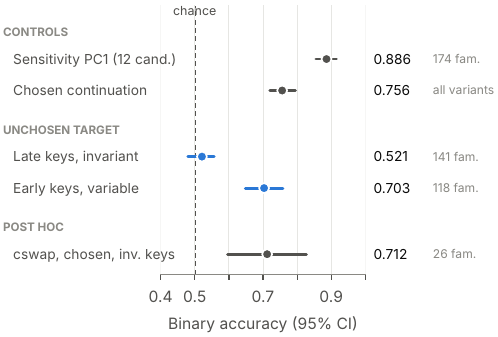}
\caption{Exploratory H1c accuracy on EVAL families (DEVX stage), with 95\% intervals. The two controls passed; in the late-key condition, where candidate keys are bitwise invariant, unchosen-continuation decoding stayed near chance (sign-flip $p = 0.18$). Late and early key conditions differ in available contextual information and populations; they are not a matched mechanistic comparison. The cswap row is post hoc on a restricted subset.}
\label{fig:h1c}
\end{figure}

The late condition better isolates a query-specific contribution, but cannot characterize probe behavior under every presentation order. We therefore examined, after the planned analyses, pairs whose keys could change and swaps affecting the chosen branch. Their post hoc status matters as much as their numerical values: they generate explanations and new hypotheses without becoming confirmatory tests retroactively.

\subsubsection{Post hoc diagnostics: query--key interaction and possible asymmetry}\label{sec:6.4.5}

For early pairs, a post hoc diagnostic compared the pointer (0.703) against a key-only decoder (0.445), a linear chosen-action + keys control (0.487), and a bilinear mean-query-by-choice control (0.517). The pointer\textquotesingle s advantage over the strongest control was +0.186 {[}0.110, 0.263{]}. This indicates a contribution from the interaction between the pre-answer query and contextual keys under this condition; it does not isolate information carried by the query alone. Moreover, the controls were trained on a mixture of variants, and their capacity specifically on early pairs was not assessed separately.

In the cswap\_L baseline, where the chosen branch\textquotesingle s continuation is swapped, the decoder achieved 0.770 across all pairs. Among only the 26 families in which both chosen candidate keys were identical across pair members, it achieved 0.712 {[}0.596, 0.827{]}, while query-free controls remained exactly at 0.50. A comparison within 25 shared families estimated a chosen-minus-unchosen advantage of +0.26 {[}0.12, 0.40{]}; however, this observation does not establish a general asymmetry. Late ordering was enforced only for the unchosen branch; the chosen subset is small, selected post hoc, and subject to different admissibility conditions. The two decoders also target different outcomes.

\subsubsection{H1c status and closure decision}\label{sec:6.4.6}

At the end of the analyses, H1c was closed with the status ``EXPLORATORY COMPLETE --- UNCHOSEN CONTINUATION NOT DETECTED.'' An independent TEST of approximately 600 families would have had limited power (around 0.35 if the true effect were close to the DEV advantage of 0.02), while the available evidence would still be exploratory. The scientific decision was not to create that TEST, not to adapt the probe once more, and to retain the chosen/unchosen asymmetry as a separate hypothesis generated post hoc and not tested confirmatorily.

Taken separately, H1, H1b, and H1c might appear to tell three different stories. Together, they describe the successive requirements for a credible inference about discarded possibilities. A correct answer does not guarantee an accessible representation; a failing probe may be insensitive; a successful probe may reconstruct its target; and a statistical difference is informative only when the null model and counterfactual worlds are appropriate. The discussion returns to these distinctions rather than to the simplistic opposition between ``the model plans'' and ``the model does not plan.''

\section{Discussion}\label{sec:7}

\subsection{Behavioral competence does not guarantee relational decodability}\label{sec:7.1}

H1 establishes an operational dissociation: the model reaches the competence required to select between the two actions, but the frozen H1 decoder fails to outperform the baselines on u2. H1b then shows that static keys make the probe insensitive to label identities; it would therefore be mistaken to read H1 NOT MET as proof that no internal representation exists. H1c improves sensitivity controls and pair construction: an instrument capable of decoding c1 and the chosen continuation shows no detectable additional signal for the unchosen continuation under the late conditions studied. This latter finding remains exploratory and limited to the decoder architecture used.

\subsection{B3: the text-and-observed-choice baseline and possible shortcuts}\label{sec:7.2}

B3 receives no internal activations and does not explicitly trace paths across multiple edges; it uses seven features of the input text together with the model\textquotesingle s observed A/B choice. Its paired accuracy of 0.195 on TEST, compared with 0.020 for the probe, warrants a restrictive interpretation of the endpoint: even an intermediate target intended to resist trivial reward lookup retains useful textual and choice-conditioned cues. Comparison with chance alone would have been misleading. The baseline must therefore be described as text-and-observed-choice, not prompt-only, and its use cannot be treated as a test against all possible graph-solving algorithms.

\subsection{Overfitting, sample size, and label representations}\label{sec:7.3}

The original H1 probe trains on 2,625 admissible worlds from 226 families and associates a 5,376-dimensional activation space with arbitrary labels that sometimes span multiple tokens. H1b reveals a limitation specific to static keys, while contextual probes can also overfit substantially. In H1c, PC1 rose from 0.394 to 0.621 and then 0.752 as training increased from approximately 218 to 421 and then 852 worlds; DEVX further raised PC1 to 0.886 on previously unseen EVAL families. This progression supports the instrumental conclusion ``sensitive for the specified controls,'' but it does not guarantee the power or adequacy of the probe for every possible type of unchosen-continuation content.

\subsection{Backward reconstruction and the absence of evidence about computation direction}\label{sec:7.4}

The P0 control shows that a solver can exploit terminal rewards to reconstruct transitions backward. Even a positive result for an ordered representation would therefore not suffice to establish a computation built ``forward.'' In this study, H1 was not met and the conditional L2 criteria were not triggered. The analogy to biological forward sweeps remains a conceptual motivation and a research program, not a demonstrated mechanism in this model. Transformer layers are also not, by themselves, a neural timescale.

\subsection{The informational value of a preregistered negative result}\label{sec:7.5}

The H1 confirmatory evaluation was performed once on LOCKED-TEST after predictors and thresholds had been frozen. On 293 admissible pairs, three contrasts failed: the NOT MET verdict cannot be retrospectively reframed as success. H1b and H1c were authorized separately and used explicitly exploratory DEV datasets, without reusing the consumed TEST. Separating these epistemic statuses is essential: post hoc analysis can guide instrument criticism and the formulation of a new hypothesis, but cannot confer confirmatory validation on that hypothesis.

Early descriptive H1 observations concerning the chosen branch (B7, c3) motivated diagnostics but did not establish an asymmetry. By the end of H1c, the chosen--unchosen difference became a more precise but still post hoc hypothesis: the chosen-side advantage in the invariant-key subset concerns only 26 families, whereas the late unchosen result concerns 141 families. A dedicated new study would require strictly comparable pairs, enforced late ordering on both branches, symmetrical positive controls, and a fresh TEST. No such study was initiated.

\subsection{Three information sources that must be separated}\label{sec:7.6}

Successive analyses led us to distinguish three questions that are often conflated: is information decodable with a given probe; is it already present in the contextual representations of the candidates; and does the pre-answer query provide additional content when the keys are held invariant? H1b accuracy with contextual keys partially answers the first question but fails to isolate the third. H1c uses bitwise key invariance between worlds to block that competing explanation. The absence of a signal in late pairs and its presence in early pairs point toward a context-dependent interaction; they do not locate a sequential prospective computation.

\subsection{Identifiability precedes statistical power}\label{sec:7.7}

Adding more families cannot fix a target that is fully reconstructible from a competing variable. H1b could not distinguish u2 from an elimination rule based on c2; H1c addresses that weakness with a dead-end swap and invariant candidate keys. But a more discriminating design also requires the correct statistical null: the initial H1c paired rule yielded 100\% false positives under a plausible dependency in predictions. Randomizing orientation and testing family means with sign flips was a substantive correction, not a mere threshold adjustment.

The invariant-comparator result in Section~\ref{sec:4.5.1} makes this distinction exact for a restricted class: if the observed choice and candidate keys are identical in the two worlds but the correct target flips, a comparator using only those values must repeat its prediction and cannot exceed 0.50 world-level accuracy on balanced pairs. This does not identify the computation in the pre-answer query, because the full prompt differs. The benefit is exclusion of a named alternative explanation by design, not a claim that all other explanations have been eliminated.

The family-level sign-flip rule addresses a separate issue: dependence between predictions. The naive paired 0.25 null led to invalid rejection behavior under a simulated anticorrelated strategy. Family-level sign flips calibrated the tested null for the H1c design; they do not justify transferring statistical conclusions from the DEVX exploratory population to a hypothetical confirmatory TEST.

\subsection{Scope of the asymmetry hypothesis}\label{sec:7.8}

The chosen continuation is decodable at accuracy 0.756 in DEVX, and the invariant-key cswap subset yields 0.712; the unchosen continuation in late pairs remains at 0.521. This contrast suggests that the model may make information about the selected branch more accessible. It would nonetheless be premature to infer an intrinsic asymmetry in its decision mechanism: differing ordering and selection conditions, few comparable pairs, decoder-head effects, and the possibility of backward reconstruction leave this interpretation hypothetical.

\subsection{Conceptual implications: prospection and intentional agency}\label{sec:7.9}

A forward sweep is a temporal and representational phenomenon documented in animals; our probes observe transformer activations at selected positions, not a temporal succession of imagined futures. The data reported here do not measure the direction in which trajectories are constructed, causal interventions, or VTE behavior. They therefore support no inference about intentional agency, subjective experience, or a system\textquotesingle s own welfare. Their contribution is to specify the conditions under which such hypotheses could eventually be examined without conflating correct choice, representation of alternatives, and the mechanism of prospection.

The conceptual thread thus returns to our starting question. We observe a choice and want to know whether discarded possibilities were represented before it was produced. Our design shows why this is harder than decoding a label: we must determine whether the label appears in the prompt, whether the probe can read it, whether another cue suffices to recover it, and whether the signal varies with the branch actually under investigation. The chosen--unchosen difference motivates future research; it is not an established general property of the agent.

These results should be read at two levels: they describe performance observed in our experiments, and they assess the strength of the inferences that such performance could support. Keeping these levels separate is itself a requirement of validity: better experimental controls do not turn a nonsignificant finding into evidence of absence, or an exploratory observation into confirmation.

\subsection{Threats to validity}\label{sec:7.10}

\subsubsection{Construct validity}\label{sec:7.10.1}

In H1 and H1b, u2 denotes a depth-2 intermediate node on an unchosen branch; in H1c, it denotes the rewarded continuation among two successors of the same branch, with symmetric dead ends. These operations directly measure neither planning, nor the computational direction of representation construction, nor subjective experience. Choice is defined by forced A/B logits, and admissibility conditions the analyses on correct model decisions. Even H1c pairs with identical keys allow a structural summary, backward propagation from the reward, or another nonprospective computation.

\subsubsection{Internal validity and observation selection}\label{sec:7.10.2}

Dataset families and node identifiers are disjoint across DEV and TEST; weights, hyperparameters, and normalization statistics are frozen before confirmatory release. Candidates are canonically ordered, access to TEST is logged, and only one evaluation is permitted. B2 is trained on permuted targets but evaluated against the true u2, with the maximum of its three versions recomputed in each bootstrap. Despite these safeguards, conditioning on correct model responses selects a subpopulation (1,000 incorrect choices excluded); the result does not generalize to worlds in which the model chooses incorrectly. Decoys cannot be decoded with the frozen six-class heads.

\subsubsection{Statistical validity and power}\label{sec:7.10.3}

For H1, DEV validation uses 77 pairs, but independent confirmation uses 293 pairs from 200 families, beyond the predefined sufficiency thresholds. All four contrasts must hold simultaneously, and the bootstrap resamples families. In H1c, independence between paired predictions cannot be assumed: the 0.25 chance threshold was invalidated under anticorrelated preferences and replaced before capture by a sign-flip test on family means. Instrument adequacy was evaluated on EVAL after splitting TRAIN/SELECT/EVAL, but the H1c u2 result remains exploratory. With the observed DEV advantage of only about 0.02, a TEST with 600 families would have had estimated power near 0.35; not running it avoids presenting a preregistered but weakly sensitive result as a general refutation.

\subsubsection{Traceability, reproducibility, and consumption of TEST}\label{sec:7.10.4}

Scripts, protocols, authorizations, probe weights, SHA-256 hashes, capture logs, thresholds, and decisions were retained by stage. H1 LOCKED-TEST is consumed; H1b v1.0 was prepared but suspended before any new TEST; H1c used DEV and DEVX captures only. Historical archives preserve H1, H1b, and H1c results without overwriting: the H1 v0.3 bundle contains 66 artifacts, and the H1b--H1c archive contains 165 read-only files (manifest hash 4724d316\ldots). Large activation tensors and data pools are referenced by hash rather than duplicated. Cumulative reported GPU/cloud cost was approximately US\$4.235, below the approved US\$6 cap. These procedures support the traceability of conclusions; they do not constitute biological evidence of prospection.

\subsubsection{Generalizability, distinct targets, and post hoc status}\label{sec:7.10.5}

The design concerns a single frozen model, one family of symbolic two-action tasks, and low-rank bilinear probes. H1 and H1c have different formats, and their results must not be compared as though drawn from one population. Competence in H1c remains moderate (approximately 0.72--0.73), excluding roughly 27\% of worlds from some analyses. The observed EVAL accuracies come from families never used for training but not from a sealed TEST; they remain exploratory. Early-pair and cswap analyses are post hoc, and the invariant-key chosen condition includes only 26 families. Successful decoding on a positive control does not imply that the probe would necessarily detect every possible form of unchosen-continuation content.

\section{Conclusion}\label{sec:8}

When a model answers A rather than B, that answer tells us which action it selects, but not whether an accessible representation of path B contributed to preparing the decision. Before the Choice grew out of this distinction between a visible decision and unrealized possibilities. The analogy with hippocampal forward sweeps supplied a question rather than an answer: do activations preceding a decision contain information about the continuation of a branch that will ultimately be rejected?

Our first experimental answer, H1 v0.3, was confirmatory and negative: the preregistered static-key probe achieved 0.020 paired accuracy on 293 LOCKED-TEST pairs, compared with 0.195 for the text-and-observed-choice baseline B3; H1 was NOT MET. This verdict concerns a specific instrument rather than the general absence of representations. Exploratory H1b revealed that static keys were insensitive to label identity, followed by an identifiability problem: u2 could be reconstructed by eliminating the chosen candidate. H1c introduced matched worlds with symmetric dead ends and invariant arrival keys, together with a calibrated family-level inference rule. Once specified controls passed (PC1 = 0.886; chosen continuation = 0.756), the late unchosen continuation was not detected (0.521; p = 0.18), while the early condition produced exploratory decoding (0.703). H1c was not evaluated on a newly sealed TEST; the chosen/unchosen asymmetry remains post hoc and unconfirmed.

The contribution of this series is not limited to an unsuccessful search for evidence of prospection. Through documented failures and controlled corrections, it identifies a set of requirements for investigating internal representations: instrument sensitivity, controls for reconstruction from prompt text and the chosen branch, verified invariance of comparison inputs, appropriate statistical calibration under dependence, and explicit separation of confirmation from exploration.

These results demonstrate neither that a model represents possible futures sequentially nor that it never does so. They do not directly address VTE, intentional agency, sentience, or AI welfare. The H1 series is closed; H2 and H3 were not tested. What remains open is not only a hypothesis about the system but also a requirement for future methods: to show which content is present, where it becomes available, how it was constructed, and whether the model actually uses it.

\subsection{Artifact availability and reproducibility}

The study reports protocol identifiers, hashes, capture integrity checks, model revision, selection rules, and results. The environment generator, independent oracle, capture and analysis code, trained DEV probe weights, protocol requests, decision records, and stage reports are publicly available at \url{https://github.com/besionisrael/before-choice} under the MIT License. Cryptographic hashes verify identity of obtained artifacts; they do not substitute for artifact access. No new experiment was performed in preparing this manuscript revision.



\bibliography{references}
\bibliographystyle{icml2027}

\newpage
\appendix
\onecolumn

\section{Fixed protocol parameters}\label{app:A}
\begin{table}[h]
\centering
\caption{Preserved protocol parameters and decisions.}
\label{tab:11}
\small
\begin{tabular}{lp{0.62\textwidth}}
\toprule
\textbf{Item} & \textbf{Fixed value / status} \\
\midrule
Model / revision & \texttt{google/gemma-4-31B-it}, revision \newline \texttt{842da3794eaa0b77d5f08bae87a17459d91ff475} \\
Primary target & Unchosen u2; depth-4 upath pairs, both members correct \\
PROBE-DEV / LOCKED-TEST & 4,800 worlds each, 300 independent families \\
Representations & $62 \times 5{,}376$ bf16 at last pre-answer token; DEV-selected $\ell^* = 32$ \\
DEV-selected probe & rank 8; regularization 0.1; 5 epochs \\
Threshold baselines & B1, B2, B3, B6 --- each must be surpassed \\
Statistic & 10,000 family bootstraps; seed 0; 95\% interval; IUT \\
Sufficiency & At least 180 depth-4 upath pairs and 100 contributing families \\
H1 / H2 / H3 status & H1 NOT MET; no L1/L2 attribution; H2 and H3 NOT TESTED \\
TEST status & TEST evaluated once and consumed; activations and oracles sealed, accesses logged \\
\bottomrule
\end{tabular}
\end{table}

\section{Artifacts and traceability}\label{app:B}

The scientific closure on November 14, 2026, is documented in BTC-H1\_FINAL\_REPORT.md, BTC-H1-TEST-EVAL\_RESULTS.md, and the archive outputs/archive/BTC-H1-v0.3/ARCHIVE\_MANIFEST.json (reported hash prefix 13f453aa\ldots). Sixty-six files were copied as read-only: protocol, decision log, scripts, DEV selection, fit registry, 16 probe weight files, normalization, authorizations, logs, and outputs from the single TEST. The two large activation datasets (DEV and TEST, approximately 3.4 GB each) are not duplicated in the archive, but their manifests and hashes are referenced. TEST pool access logs (four entries) and activation-access logs (39 analytical reads) were frozen as snapshots. Independent reproduction requires the original datasets and activation files in addition to the small-artifact archive.

Reference hashes (SHA-256):
\begin{itemize}\small
\item LOCKED-TEST-D5 seal:\\ \texttt{f1d541e297a261ab13573e6f36ea4082847d4fd968cef7a228559f86d0918392}
\item LOCKED-TEST-D3-ACT seal:\\ \texttt{fe7c57569a727eaf9b75bbbd0d01be1b917e3b50495f54a2fb2dcec972980597}
\item PROBE-DEV activation manifest:\\ \texttt{306573de935f2092b49a60171b6831443a52f4f9132649949e0e5c1aef3d4d8d}
\item TEST result EVAL\_RESULT.json hash:\\ \texttt{e021937853ee2919c85b79cc1c804ae5967ffae3a848fdf28e7a94b49f3c58b2}
\item per\_world.jsonl hash:\\ \texttt{5f3aa6de2b872a66b6cd2fdbe06ccf022def789c368914fb531434a8176dab33}
\end{itemize}
These SHA-256 values make artifact identity verifiable but do not replace access to the artifacts or independent verification of procedures.

\section{Decision chronology}\label{app:C}

H1 v0.3 (confirmatory NOT MET): BTC-H1\_FINAL\_REPORT.md and BTC-H1-TEST-EVAL\_RESULTS.md. H1b (exploratory): BTC-H1b-EXPLORATORY\_RESULTS.md and BTC-H1b-X3X4\_RESULTS.md; the identifiability decision and suspended TEST are documented in BTC-H1b\_METHOD\_DECISION\_NOTE.md and BTC-H1b\_V1\_FREEZE\_VERIFICATION.md. H1c (exploratory; no sealed TEST): BTC-H1c\_DESIGN\_PROPOSAL.md, BTC-H1c\_PROTOCOL\_DRAFT.md, BTC-H1c\_DEV\_RESULTS.md, BTC-H1c\_INSTRUMENT\_DIAGNOSTIC.md, BTC-H1c\_DEVX\_RESULTS.md, and BTC-H1c\_FINAL\_DIAGNOSTIC.md. Overall closure and artifact manifest: BTC\_H1\_RESEARCH\_CONSOLIDATION.md and outputs/archive/BTC-H1b-H1c/ARCHIVE\_MANIFEST.json. This manuscript revision does not alter historical protocols or results; H2 and H3 were not tested.

\end{document}